\documentclass[usenatbib,useAMS,twocolumn]{mn2e}
\usepackage{psfig}
\usepackage{ltxtable}
\usepackage[latin1]{inputenc}
\newcommand{\dd}[2]{\frac{{\rm d} #1}{{\rm d} #2}}
\newcommand{\Dd}[2]{\frac{{\rm D} #1}{{\rm D} #2}}
\newcommand{\dpart}[2]{\frac{\partial #1}{\partial #2}}
\newcommand{\dth}[3]{(\dpart{#1}{#2})_{#3}}

\font\zzz=cmmib10

\def\pmb#1{\setbox0=\hbox{#1}
  \kern-.025em\copy0\kern-\wd0
  \kern.05em\copy0\kern-\wd0
  \kern-.025em\raise.0433em\box0 }

\def\bnabla{\pmb{$\nabla$}}
\def\bmu{\hbox{\zzz\char'026}}          

   \title[Formalism for the convective Urca process]
{A two-stream formalism for the convective Urca process}

   \author[P. Lesaffre {\it et al.}]
{P. Lesaffre,$^{1,2}$\thanks{lesaffre@ast.cam.ac.uk}      
        Ph.\ Podsiadlowski$^{2}$
	and C. A. Tout $^{1}$\\
$^{1}$Institute of Astronomy, Madingley Road, Cambridge 
CB3~0HA, UK \\
$^{2}$University of Oxford, Department of Astrophysics,
 Oxford OX1~3RH, UK}

\begin{document}

   \date{Received September 15, 1996; Accepted March 16, 1997}

   \maketitle

\begin{abstract} 
%

   We derive a new formalism for convective motions involving two
   radial flows. This formalism provides a framework for convective
   models that guarantees consistency for the chemistry and the
   energy budget in the flows, allows time-dependence and accounts for
   the interaction of the convective motions with the global
   contraction or expansion of the star.  In the one-stream limit the
   formalism reproduces several existing convective models and allows
   them to treat the chemistry in the flows.  We suggest a version of
   the formalism that can be implemented easily in a stellar evolution
   code.

   We then apply the formalism to convective Urca cores in
   Chandrasekhar mass white dwarfs and compare it to previous
   studies. We demonstrate that, in degenerate matter, nuclear
   reactions that change the number of electrons strongly influence
   the convective velocities and we show that the net energy budget is
   sensitive to the mixing. We illustrate our model by computing {\it
   stationary} convective cores with Urca nuclei. Even a very small
   mass fraction of Urca nuclei (as little as  $10^{-8}$) strongly influences
   the convective velocities. We conclude that the proper modelling of
   the Urca process is essential for determining the ignition
   conditions for the thermonuclear runaway in Chandrasekhar-mass
   white dwarfs.
\end{abstract}

\section{Introduction}

\citet{GS41} were the first to recognise the Urca process (electron
captures and emissions on pairs of nuclei that can be converted into
each other by electron captures/beta decays) as a potentially strong
source of neutrino cooling in degenerate stars.  This process is
already responsible for significant cooling during the late radiative
phase of accreting C/O white dwarfs. For each Urca pair, the cooling
occurs at a mass shell, a so-called Urca shell, determined by the
characteristic density for the pair at which the electron
captures/beta decays take place.  When carbon burning starts, a
convective core grows and soon engulfs the Urca shells.  The
convective motions across the Urca shells back and forth directly
affect the net energy release as well as the net amount of electrons
captured. The resulting phenomenon is called the convective Urca
process and is a key ingredient in linking the late evolution of the
progenitor of a Type Ia supernova (SN~Ia) with the subsequent
explosion \citep{P72}.

Over the last 30 years there have been numerous studies of the convective Urca
process with often mutually exclusive conclusions.  \citet{B73}
realised that nuclear heating in Urca matter outside chemical
equilibrium dominates over the neutrino losses.  On the other hand,
\citet{CA75} stressed the cooling effect of the work done by
convection.  In the most detailed work to date, \citet{I78a,I78b,I82}
computed the evolution of an accreting white dwarf including the
detailed chemistry of many Urca pairs. He realised that the turn-over
time scales would be of the same order as the chemical time scales for
the Urca reactions. This, he concluded, implied that the mixing
processes caused by the growth of the convective core would affect the
heating/cooling by the Urca process.  However numerical problems
caused by his treatment of the mixing prevented him from following
the computations up to the thermal runaway.

 \citet{BW90} revisited the problem of the convective Urca process and
 provided a clear summary of the convective Urca mechanism though
 later \citet{M96} and \citet{SBW99} pointed out a mistake in their
 treatment and argued that more attention needs to be paid to the
 kinetic energy of convection.  \citet{B01} showed that the
 feedback of the Urca process on convection itself should be taken
 into account. As this summary shows, a consistent picture of
 convection that properly treats the chemistry is still missing and is
 badly needed in order to address the problem of the convective Urca
 process.

 Such a theory for chemistry coupled with convection was attempted by
 \citet{E83} with the help of a ``simple rule-of-thumb procedure''.
 Later, \citet{G93} produced a physically consistent model of
 convection which includes chemistry based on a statistical approach for the
 convective blobs. However, their model did not necessarily conserve
 energy and has only been checked without chemistry.

 Here we devise a model for convection that ensures
 energy conservation. We start with the conservation equations of
 radiation hydrodynamics in a spherical configuration. Using a simple
 geometry, which mimics convective rising plumes, we derive a
 physically self-consistent model of convection which includes
 time-dependent chemistry. We then compare this formalism to
 previously derived models of convection. In the process we obtain a
 model for the convective Urca process that addresses all the problems
 mentioned above, nuclear heating, mixing, convective work, kinetic
 energy and the feedback of the Urca process on convection. To
 illustrate the formalism, we apply it to stationary convective Urca
 regions and show how it affects the energy budget, the convective
 properties and the chemical stratification. In a follow-up paper, we
 plan to apply the formalism in a realistic, time-dependent stellar
 model, coupled with a complete nuclear reaction network, to determine
 the ignition conditions for the thermonuclear runaway in a SN Ia.

 In section 2 we derive the
 basic equations for the two-stream formalism and suggest a simple
 model for the exchange of matter, momentum and energy between rising
 and descending flows. In section 3 we compare our model to existing
 models of convection.  In section 4 we describe the convective Urca
 process in view of our model. In section 5 we compute stationary Urca
 convective cores. We discuss and summarise our results in sections 6
 and 7.

\section{A two-stream formalism for convection}

 \citet{C93} designed a two-stream algorithm to post-process the
 evolution of chemical species in convective regions. This model was
 well suited to the study of convective regions in which the chemical
 time scales were shorter than the convective turn-over time
 scales. But, as a post-processing algorithm, it did not tackle the
 feedback effects the chemistry could have on the convection. Here we
 extend his ideas to all state variables in the two streams and
 explore the interactions between mass, energy, momentum and chemical
 transfers between the streams.

\begin{figure}
\centerline{
\psfig{file=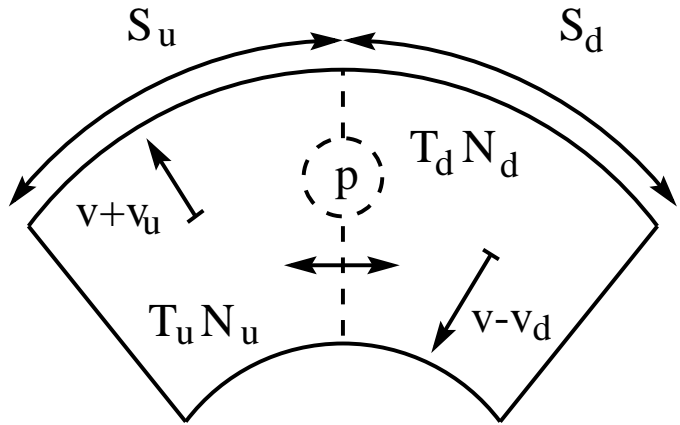,angle=-90,width=9cm}
}
\caption{Geometrical configuration of the two streams.}
\label{sketch}
\end{figure}

  Let us consider a sphere of gas with purely radial velocities and
  stratified properties. We assume that, on a shell of radius $r$,
  there are two different velocities associated with the two streams.
  We further assume that, on this shell, all gas parcels moving with a
  given velocity have the same state, they have homogeneous
  temperature, pressure and chemical composition. Figure
  (\ref{sketch}) schematically shows the geometrical configuration we
  have in mind.

  One of the two velocities has to be greater than the other, and we
  refer to the gas moving with this velocity as the upward moving gas,
  even though both velocities could be negative in principle. We write
  the two velocities as $v+v_u$ and $v-v_{\rm d}$, where $v_u$ and
  $v_{\rm d}$ are both positive and are associated with the upward and
  the downward moving fluid, respectively. The ambient velocity $v$
  will be defined further below.

  We now affix suffixes `u' to upward moving fluid properties and
  `d' to downward moving fluid properties.  We assume that the
  sound-crossing time of a horizontal section of one stream is short
  compared to any other time scale and hence make the approximation
  that the pressure $p=p_{\rm u}=p_{\rm d}$ is the same in the upward
  and downward moving fluids. The state of the gas in both fluids is
  now completely determined by the temperatures $T_{\rm u}$ and
  $T_{\rm d}$ and abundance vectors ${\bf N}_{\rm u}$ and ${\bf
  N}_{\rm d}$ (number of particles per unit mass), provided we know
  the equation of state of the gas. The latter equation provides the
  mass densities $\rho_{\rm u}(p,T_{\rm u},{\bf N}_{\rm u})$ and
  $\rho_{\rm d}(p,T_{\rm d},{\bf N}_{\rm d})$, and the specific
  energies $e_{\rm u}$ and $e_{\rm d}$, along with the radiative
  volumic energies $E_{\rm u}$ and $E_{\rm d}$.

  We define $S_{\rm u}$ and $S_{\rm d}$ as the areas of the
  surfaces occupied by the two fluids at the shell of radius
  $r$. Therefore
\begin{equation}
\label{S}
 S=S_{\rm u}+S_{\rm d}=4 \pi  r^2 
\end{equation}
  and we define the velocity $v$ by setting the mass flow
\begin{equation}
\label{dotm}
\dot{m}=S_{\rm u} v_{\rm u} \rho_{\rm u}=S_{\rm d} v_{\rm d} \rho_{\rm d}
\mbox{.} \end{equation}
  This defines $v$ as the radial velocity of the centre of a mass
  shell. The net mass flow through the shell moving at velocity $v$ is
  zero.

 We now write the equations for the variation of mass, momentum and
 energy for each of the two fluids and define the exchange terms. We
 then compute the equations for the average fluid and the specific
 equations of evolution for both fluids, before computing the
 differential evolution between the two fluids. Finally, we propose a
 very simple model for the exchange terms between the streams.

\subsection{Conservation equations}
We consider the mass, momentum and energy on a shell at radius $r$ for
each of the two fluids. This allows us to define the exchange
terms in a conservative way and makes it easier to derive the
mean equations. Moreover, with this approach the specific exchange
terms can be defined more rigorously.

The viscosity of the fluid is neglected as well as the molecular
diffusion.  All horizontal effects are implicitly included in the
exchange terms.

  \subsubsection {Mass}

 To simplify the derivation, we neglect the mass changes due to the
 nuclear transformations and assume that mass is perfectly conserved.
 We treat the corresponding nuclear energy production only in the
 energy equation. The rate of change of mass is hence equal to the sum
 of the mass flux in the radial direction and sideways,

\begin{equation}
\label{massu}
\dpart{}{t}(S_{\rm u} \rho_{\rm u})=-\dpart{}{r}[S_{\rm u} \rho_{\rm u} (v+v_{\rm u})]+\dot{M}
\mbox{,} \end{equation}
and
\begin{equation}
\dpart{}{t}(S_{\rm d} \rho_{\rm d})=-\dpart{}{r}[S_{\rm d} \rho_{\rm d} (v-v_{\rm d})]-\dot{M}
\label{massd}
\mbox{,} \end{equation} where $\dot{M}$ represents the mass per unit
radius and unit time exchanged in the shell from the downward to the
upward moving fluid.

\subsubsection{Momentum}
The rate of change of momentum is the sum of the momentum flux 
(ram and thermal pressure) in the radial direction and sideways, added
to the gravitational forces where the gravitational potential is
assumed to be spherical,

\begin{equation}
\label{momentumu}
\dpart{}{t}[S_{\rm u} \rho_{\rm u} (v+v_{\rm u})]=-S_{\rm u}\dpart{}{r}[\rho_{\rm u} (v+v_{\rm u})^2+p]-
S_{\rm u}\rho_{\rm u}\frac{Gm}{r^2}+\dot{Q}
\mbox{,} \end{equation}
and
\begin{equation}
\dpart{}{t}[S_{\rm d} \rho_{\rm d} (v-v_{\rm d})]=-S_{\rm d}\dpart{}{r}[\rho_{\rm d} (v-v_{\rm d})^2+p]-
S_{\rm d}\rho_{\rm d}\frac{Gm}{r^2}-\dot{Q}
\label{momentumd}
\mbox{.} \end{equation}
 Here $G$ is Newton's constant of gravitation, $m$ is the total
 mass contained inside the sphere of radius $r$,
\begin{equation}
m=\int_0^{r} (S_{\rm u}\rho_{\rm u}+S_{\rm d}\rho_{\rm d})\,{\rm d}r'
\end{equation}
and $\dot{Q}$ is the momentum per unit mass and unit time
exchanged in this shell from the upward to the downward moving fluid.

\subsubsection{Energy}
 The rate of change of energy is the advective plus diffusive energy
 flux, in the radial direction and sideways, added to the work of
 pressure forces and heat generation,
\begin{displaymath}
\dpart{}{t}(S_{\rm u} \rho_{\rm u} e_{\rm u})=-\dpart{}{r}[S_{\rm u}
  \rho_{\rm u} e_{\rm u} (v+v_{\rm u})] -p\dpart{}{r}[S_{\rm
    u}(v+v_{\rm u})]
\end{displaymath}
\begin{equation}
-p\dpart{}{t}(S_{\rm u}) +S_{\rm
u} \rho_{\rm u}\left(\epsilon_{\rm u}-\frac{\partial L_{\rm u}}{S_{\rm u}
\rho_{\rm u} \partial r}\right)+\dot{E}
\label{energyu}
\end{equation}
and
\begin{displaymath}
\dpart{}{t}(S_{\rm d} \rho_{\rm d} e_{\rm d})=-\dpart{}{r}[S_{\rm d} \rho_{\rm d} e_{\rm d} (v-v_{\rm d})]
-p\dpart{}{r}[S_{\rm d}(v-v_{\rm d})]
\end{displaymath}
\begin{equation}
-p\dpart{}{t}(S_{\rm d})
+S_{\rm d} \rho_{\rm d}\left(\epsilon_{\rm d}-\frac{\partial L_{\rm d}}{S_{\rm d} \rho_{\rm d} \partial r}\right)-\dot{E}
\label{energyd}
\mbox{.} \end{equation}
Here, $L$ is the luminosity carried by the conductive and radiative processes
(in the radial direction),
\begin{eqnarray}
L_{\rm u}&=&-\frac{S_{\rm u} c}{3 \kappa_{\rm u}}\dpart{E_{\rm u}}{r}\\
&=&\rho_{\rm u} c_{\rm Pu} \chi_{\rm u} \dpart{T_{\rm u}}{r} S_{\rm u},
\end{eqnarray}
and
\begin{eqnarray}
L_{\rm d}&=&-\frac{S_{\rm d} c}{3 \kappa_{\rm d}}\dpart{E_{\rm d}}{r}\\
&=&\rho_{\rm d} c_{\rm Pd} \chi_{\rm d} \dpart{T_{\rm d}}{r} S_{\rm d}
\mbox{,} \end{eqnarray}
where $E$ is the radiative energy per unit of volume given by the
equations of state, $c$ is the speed of light, $\kappa$ is the flux
weighted total opacity for the conductive and radiative processes (in the
diffusion approximation), $\chi$ is the thermal diffusion
coefficient and $c_{\rm P}$ is the heat capacity at constant
pressure.\\ 
The net energy production per unit time and per unit mass
is $\epsilon$.  This can include the nuclear energy production if the
rest mass energy is not yet included in the specific energy, and it
includes the neutrino losses.
The energy per unit time and per unit radius exchanged in the shell
from the upward to the downward moving fluid is $\dot{E}$.

\subsubsection{Chemistry}
The rate of change of a species is its flux in the radial direction and
sideways added to its chemical rate of change,

\begin{equation}
\label{chemistryu}
\dpart{}{t}(S_{\rm u} \rho_{\rm u} {\bf N}_{\rm u})=-\dpart{}{r}[S_{\rm u} \rho_{\rm u} {\bf N}_{\rm u} (v+v_{\rm u})]
+S_{\rm u} \rho_{\rm u} {\bf R}_{\rm u}+\dot{{\bf N}}
\end{equation}
and
\begin{equation}
\dpart{}{t}(S_{\rm d} \rho_{\rm d} {\bf N}_{\rm d})=-\dpart{}{r}[S_{\rm d} \rho_{\rm d} {\bf N}_{\rm d} (v-v_{\rm d})]
+S_{\rm d} \rho_{\rm d} {\bf R}_{\rm d}-\dot{{\bf N}}
\label{chemistryd}
\mbox{,} \end{equation}
where $\dot{N}_j$ represents the mass of species $j$ per unit radius
and per unit time exchanged in the shell from the downward moving
fluid to the upward moving fluid, and ${\bf R}_j$ is the rate
of change of species $j$ per unit time and unit mass due to nuclear
reactions.

\subsection{Equations for the mean fluid}
\label{mean}

  When we sum up each pair of the above equations we obtain the equations
for the variation of the mean mass, energy and momentum. We first derive
these equations in their volumic form and then gather them all in 
a more familiar specific form.

\subsubsection{Mass}
  The mean density $\rho$ is defined by
\begin{equation}
\rho S={\rho_{\rm u} S_{\rm u} +\rho_{\rm d} S_{\rm d}}
\mbox{.} \end{equation}
  The equation for mass conservation then reads
\begin{equation}
\label{mmass}
\dpart{}{t}(S \rho)=-\dpart{}{r}(S \rho v),
\end{equation}
which is the unchanged continuity equation for the mean fluid.

\subsubsection{Momentum}

  The mean equation for momentum is more complex and involves several 
additional terms,
\begin{equation}
\label{mmomentum}
\dpart{}{t}(S \rho v)=-S\dpart{}{r}(\rho v^2+p)-
S\rho\frac{Gm}{r^2}+S\rho X_{\rm conv}
\mbox{,} \end{equation}
where
\begin{equation}
S\rho X_{\rm conv}=X_1+X_2+X_3
\mbox{,} \end{equation}
\begin{equation}
X_1=-S_{\rm u}\dpart{}{r}(\rho_{\rm u} v_{\rm u}^2)-S_{\rm d}\dpart{}{r}(\rho_{\rm d} v_{\rm d}^2)
\mbox{,} \end{equation}
\begin{equation}
X_2=-2v[S_{\rm u}\dpart{}{r}(\rho_{\rm u} v_{\rm u})-S_{\rm d}\dpart{}{r}(\rho_{\rm d} v_{\rm d})]
\mbox{ and} \end{equation}
\begin{equation}
X_3=(\rho_{\rm d}-\rho_{\rm u})v^2\left[\frac{S_{\rm u}}{S}\dpart{S_{\rm d}}{r}-\frac{S_{\rm d}}{S}\dpart{S_{\rm u}}{r}\right]
\mbox{,} \end{equation}
where we have used equation (\ref{S}).
Usually, the evolution of the star is quasi-static, which means that $v$ is 
negligible. Then only $X_1$ should be retained because $X_2$ and $X_3$ are
of order 1 and 2 in $v$.

$X_1$ accounts for a convective pressure support
in the flow. It slightly changes the hydrostatic equilibrium but is
usually negligible in the subsonic regime.

\subsubsection{Energy}

We define the mean specific energy as
\begin{equation}
S\rho e =\rho_{\rm u} S_{\rm u} e_{\rm u} + \rho_{\rm d} S_{\rm d} e_{\rm d}
\end{equation}
and the mean energy generation per unit mass $\epsilon$ in
the same way as
\begin{equation}
\rho \epsilon =\rho_{\rm u} S_{\rm u} \epsilon_{\rm u} + \rho_{\rm d} S_{\rm d} \epsilon_{\rm d}.
\end{equation}
The radiative and conductive luminosity is given by
\begin{equation}
L=L_{\rm u}+L_{\rm d}.
\end{equation}
  Then the equation for the mean energy is rather simple and
  includes only two additional terms,
\begin{displaymath}
\dpart{}{t}(S \rho e)=-\dpart{}{r}(S \rho e v)
-p\dpart{}{r}(Sv)
\end{displaymath}
\begin{equation}
\label{menergy}
+S \rho\left(\epsilon-\frac{\partial L+L_{\rm conv}}{S \rho \partial r}\right)
+S \rho W_{\rm conv}
\mbox{,} \end{equation}
where
\begin{eqnarray}
L_{\rm conv}&=&\dot{m}(h_{\rm u}-h_{\rm d})=\frac12 S\rho u(h_{\rm u}-h_{\rm d}), \\
W_{\rm conv}&=&\frac12u(\frac1{\rho_{\rm u}}-\frac1{\rho_{\rm d}})\dpart{p}{r}.
\end{eqnarray}
Here $h=e+p/\rho$ is the enthalpy and $u=2\dot{m}/(S\rho)$ is the mean
convective velocity.\\ Note that the convective luminosity $L_{\rm
conv}$ naturally appears as an enthalpy flux.\\ The term $W_{\rm conv}$ can be
interpreted as the work done by the buoyancy forces. Because the
pressure is decreasing upward, and usually $\rho_{\rm u}<\rho_{\rm
d}$, it always provides a sink term for the energy.

Note that this approach cannot be
used to write down a general equation for the entropy because the
thermodynamic relations are only valid for the up and down streams
separately but not for the mean fluid.  We will derive an entropy
equation later in the one-flow limit (see section \ref{entropy}).

\subsubsection{Chemistry}

The equation for the mean abundances is
\begin{equation}
\label{mchemistry}
\dpart{}{t}(S \rho {\bf N})=
-\dpart{}{r}(S \rho {\bf N} v)
+S \rho {\bf R}-S\rho \dpart{}{m}{\bf F}_{\rm conv}
\mbox{,} \end{equation}
where
\begin{equation}
S\rho{\bf N}=S_{\rm u}\rho_{\rm u}{\bf N}_{\rm u}
+S_{\rm d}\rho_{\rm d}{\bf N}_{\rm d}
\end{equation}
and
\begin{equation}
{\bf F}_{\rm conv}=\dot{m}({\bf N}_{\rm u}-{\bf N}_{\rm d})
\mbox{,} \end{equation}
which we interpret as a diffusion flow in section \ref{model}.

\subsubsection{Mean specific equations}

  We now summarise all of the mean equations derived in the previous section
in their specific form (i.e.\ per unit mass), where we use the notation
  $\Dd{x}{t}=\dpart{x}{t}+v\dpart{x}{r}$ and ${\rm
  div}(x)=\frac1{S}\dpart{Sx}{r}$:

\begin{equation}
\label{smmass}
\frac1{\rho}\Dd{\rho}{t}=-{\rm div}(v)
\mbox{,} \end{equation}
\begin{equation}
\label{smmomentum}
\Dd{v}{t}-2\frac{v^2}{r}=-\frac1{\rho}\dpart{p}{r}-\frac{Gm}{r^2}+X_{\rm conv}
\mbox{,} \end{equation}
\begin{equation}
\label{smenergy}
\Dd{e}{t}+p\Dd{}{t}(\frac1{\rho})=\epsilon-\dpart{L}{m}
-\dpart{L_{\rm conv}}{m}+W_{\rm conv}
\end{equation}
and
\begin{equation}
\label{smchemistry}
\Dd{{\bf N}}{t}={\bf R}-\dpart{}{m}{\bf F}_{\rm conv}
\end{equation}
with
\begin{equation}
X_{\rm conv}\simeq -\frac{1}{S\rho}\left[S_{\rm
    u}\dpart{}{r}(\rho_{\rm u} v_{\rm u}^2)+S_{\rm
    d}\dpart{}{r}(\rho_{\rm d} v_{\rm d}^2)\right] \mbox{,} 
\end{equation}
\begin{equation}
L_{\rm conv}=\dot{m}(h_{\rm u}-h_{\rm d})
\mbox{,}
\end{equation}
\begin{equation}
W_{\rm conv}=\frac12u\left(\frac1{\rho_{\rm u}}-\frac1{\rho_{\rm d}}\right)\dpart{p}{r}\mbox{,}
\end{equation}
 and
\begin{equation}
{\bf F}_{\rm conv}=\dot{m}({\bf N}_{\rm u}-{\bf N}_{\rm d})
\mbox{.} 
\end{equation}

  These equations are the usual equations of radiative stellar
  evolution with additional terms due to the differential motions in the
  two fluids. The latter terms depend on the convective velocities
  $v_{\rm u}$ and $v_{\rm d}$, as well as on differences between quantities in the
  two flows. We derive equations for the mean convective
  velocities and those differences in the one flow limit in section
  \ref{dequations}.

\subsection{Specific equations}
  We first rewrite the conservation equations in their specific form. This
  will allow the derivation of the equations in the one-flow limit 
  when we take the difference between the specific equations for
  the up and down motions.



\subsubsection{Momentum}

\begin{displaymath}
\dpart{}{t}(v+v_{\rm u})+(v+v_{\rm u})\dpart{}{r}(v+v_{\rm u})
-(v+v_{\rm u})^2\frac1{S_{\rm u}}\dpart{S_{\rm u}}{r}=
\end{displaymath}
\begin{equation}
-\frac1{\rho_{\rm u}}\dpart{p}{r}
-\frac{Gm}{r^2}
+\frac1{S_{\rm u}\rho_{\rm u}}[\dot{Q}-\dot{M}(v+v_{\rm u})]
\label{smomentumu}
\end{equation}
and
\begin{displaymath}
\dpart{}{t}(v-v_{\rm d})+(v-v_{\rm d})\dpart{}{r}(v-v_{\rm d})
-(v-v_{\rm d})^2\frac1{S_{\rm d}}\dpart{S_{\rm d}}{r}=
\end{displaymath}
\begin{equation}
-\frac1{\rho_{\rm d}}\dpart{p}{r}
-\frac{Gm}{r^2}
-\frac1{S_{\rm d}\rho_{\rm d}}[\dot{Q}-\dot{M}(v-v_{\rm d})]
\label{smomentumd}
\mbox{.}
\end{equation}

\subsubsection{Energy}
\begin{displaymath}
\dpart{e_{\rm u}}{t} +(v+v_{\rm u})\dpart{e_{\rm u}}{r}
+p\left[\dpart{}{t}\frac1{\rho_{\rm u}}+(v+v_{\rm
u})\dpart{}{r}\frac1{\rho_{\rm u}}\right]=
\end{displaymath}
\begin{equation}
\label{senergyu}
\epsilon_{\rm u}
-\frac1{S_{\rm u}\rho_{\rm u}}\dpart{L_{\rm u}}{r}
+\frac1{S_{\rm u}\rho_{\rm u}}(\dot{E}-\dot{M}h_{\rm u})
\end{equation}
and
\begin{displaymath}
\dpart{e_{\rm d}}{t} +(v-v_{\rm d})\dpart{e_{\rm d}}{r}
+p\left[\dpart{}{t}\frac1{\rho_{\rm d}}+(v-v_{\rm
d})\dpart{}{r}\frac1{\rho_{\rm d}}\right]=
\end{displaymath}
\begin{equation}
\label{senergyd}
\epsilon_{\rm d}
-\frac1{S_{\rm d}\rho_{\rm d}}\dpart{L_{\rm d}}{r}
-\frac1{S_{\rm d}\rho_{\rm d}}(\dot{E}-\dot{M}h_{\rm d})\\
\mbox{.}
\end{equation}

\subsubsection{Chemistry}
\begin{equation}
\label{schemistryu}
\dpart{{\bf N}_{\rm u}}{t} +(v+v_{\rm u})\dpart{{\bf N}_{\rm u}}{r}=
{\bf R}_{\rm u}+\frac{1}{S_{\rm u} \rho_{\rm u}}({\bf
\dot{N}}-\dot{M}{\bf N}_{\rm u}) 
\end{equation}
and
\begin{equation}
\dpart{{\bf N}_{\rm d}}{t} +(v-v_{\rm d})\dpart{{\bf N}_{\rm d}}{r}=
{\bf R}_{\rm d}-\frac{1}{S_{\rm d} \rho_{\rm d}}({\bf
\dot{N}}-\dot{M}{\bf N}_{\rm d})\mbox{.}
\label{schemistryd}
\end{equation}

\subsection{The difference equations in the one stream limit}
\label{dequations}
  In the following we make the approximation that the relative
  difference of a quantity between the two streams is small. For each
  quantity $x$ we write $x_{\rm u}=x+\Delta x$ and $x_{\rm
  d}=x-\Delta x$ with $\Delta x\ll x$. We note that the previously
  defined mean convective velocity $u$ is in fact the arithmetic mean
  of $v_{\rm u}$ and $v_{\rm d}$. Hence, we write $v_{\rm u}=u+\Delta
  u$ and $v_{\rm d}=u-\Delta u$. In contrast, we keep the former
  definition for $S$ as the {\it total} surface of the shell of radius
  $r$ so that $S_{\rm u}=\frac12S+\Delta S$ and $S_{\rm
  d}=\frac12S-\Delta S$.  We then compute the difference of the
  specific equations (\ref{smomentumu})--(\ref{schemistryd}) between
  the two fluids and neglect second order terms in the $\Delta$
  quantities.  

  In this approximation, the averages defined in section \ref{mean}
  above are all arithmetic means. We also note that the standard
  thermodynamic relations hold for the $\Delta$ quantities, which
  greatly helps when deriving the entropy equation (see section
  \ref{entropy}).

 Equation (\ref{dotm}) yields
\begin{equation}
 2 \frac{\Delta S}{S}+\frac{\Delta \rho}{\rho}+\frac{\Delta u}{u}=0
\end{equation}
  and this allows us to eliminate $\Delta S$ in what follows. 
It can be thought of as the equation governing the
  convective motions with $u$ as the convective velocity.

  \subsubsection{Mass}
  Taking the difference between equations (\ref{massu}) and (\ref{massd}) and
using equation (\ref{smmass}) we find
\begin{equation}
\label{dmass}
\Dd{}{t}\left(\frac{\Delta u}{u}\right)=
\frac1{S\rho}\dpart{}{r}(S\rho u)-\frac{2\dot{M}}{S\rho}
\mbox{.} \end{equation}
 This equation describes the evolution of the asymmetry $\Delta u$
of the drift motions relative to the mean velocity $v$.

  \subsubsection{Momentum}
The difference between equations (\ref{smomentumu}) and
(\ref{smomentumd}) yields
\begin{displaymath}
\Dd{u}{t}+u\left(\dpart{(v+2\Delta u)}{r}-4\frac{v+\Delta u}{r}\right)=
\end{displaymath}
\begin{displaymath}
\frac{\Delta \rho}{\rho^2}\dpart{p}{r}
-(v^2+u^2)\dpart{}{r}\left(\frac{\Delta \rho}{\rho}\right)
\end{displaymath}
\begin{equation}
\label{dmomentum}
-v^2\dpart{}{r}\left(\frac{\Delta u}{u}\right)
+\frac2{S\rho}[\dot{Q}-\dot{M}(v+2\Delta u)]
\mbox{.\\} \end{equation}
  This equation describes the time evolution of the convective
  velocity $u$. The main source term is the acceleration due to the
  buoyancy force while the main sink term comes from the momentum exchange
  term between the two flows.\\ 
  Apart from the $\Delta u$
  terms, $-(v^2+u^2)\dpart{}{r}(\frac{\Delta \rho}{\rho})$ is the only
  non-local term. That is it is the only term that involves spatial derivatives
  of convective properties. 

  \subsubsection{Energy}
The difference between equations (\ref{senergyu}) and (\ref{senergyd}) 
yields
\begin{displaymath}
\Dd{\Delta e}{t}-p\Dd{}{t}\left(\frac{\Delta \rho}{\rho^2}\right)
+u\left(\dpart{e}{r}+p\dpart{}{r}\frac1{\rho}\right)=
\end{displaymath}
\begin{equation}
\label{denergy}
\Delta \epsilon+
2\frac{\Delta u}{u}\dpart{L}{m}+2\dpart{\Delta L}{m} 
+\frac2{S\rho}(\dot{E}-\dot{M}h)
\mbox{.\\} \end{equation}
 This equation implicitly determines the time evolution of the
 temperature difference between the two flows as we show in
 section (\ref{MLT}).  Note that the $\Delta \epsilon$ term may help to
 either increase or decrease such a temperature difference, depending on the
 temperature dependence of the energy generation rate.
 The other non-local term in the energy equation, $2\dpart{\Delta L}{m}$,
 accounts for differenital thermal diffusion along the two columns of fluid.

  \subsubsection{Chemistry}
The difference between equations (\ref{chemistryu}) and
(\ref{chemistryd}) yields
\begin{equation}
\label{dchemistry}
\Dd{\Delta {\bf N}}{t}
+u\dpart{{\bf N}}{r}=
\Delta {\bf R}
+\frac2{S\rho}(\dot{{\bf N}}-\dot{M}{\bf N})
\mbox{.\\} \end{equation}
 This equation describes the evolution of the chemical composition
 difference between the two flows. In some situations, the temperature
 dependence of the nuclear reaction rates may affect this difference through
 the $\Delta {\bf R}$ term.

\subsection{A model for the exchange terms} 
\label{model}

  All horizontal motions and transport phenomena are modelled through
  the exchange terms. These need to be specified to close the two
  systems of equations (\ref{smmass})--(\ref{smchemistry}) and
  (\ref{dmass})--(\ref{dchemistry}).  With these, the equations
  provide a complete time-dependent model for convection which
  guarantees the conservation of mass, momentum, energy and chemical
  transformations. Most of the existing convective models provide
  approximations for these exchange terms. Here we present a simple,
  somewhat {\it ad hoc}, but physically plausible choice for these
  exchange terms,

\begin{equation}
\label{exchange1}
\dot{Q}=-(v_{\rm u}+v_{\rm d})\dot{m}/\lambda+\dot{M}v
\mbox{,} \end{equation}
\begin{equation}
\label{exchange2}
\dot{E}=(h_{\rm d}-h_{\rm u})\dot{m}/\lambda +\beta\frac{
\chi}{u\lambda} c_p (T_{\rm d}-T_{\rm
u})\frac{\dot{m}}{\lambda}+\dot{M}h,
\end{equation}
and 
\begin{equation}
\dot{{\bf N}}=({\bf N}_{\rm d}-{\bf N}_{\rm u})\dot{m}/\lambda+\dot{M}{\bf N}
\label{exchange3}
\mbox{.\\} \end{equation} 
 The first terms in these expressions account
 for momentum, energy and chemical exchanges without any net transfer
 of mass. They are designed to mix the two fluids on a length scale
 $\lambda$. The second term in the energy exchange equation
 (\ref{exchange2}) accounts for horizontal heat diffusion across the
 edges of the streams. The parameter $\beta$ is a form factor which
 fixes the ratio of the perimeter of the streams to their separation
 times $S/\lambda^2$. We use $\beta=9/2$ to recover the exact
 formulation of classical mixing-length theory (MLT; \citet{B58} as
 presented in \citet{K90}). The last terms in expressions
 (\ref{exchange1})--(\ref{exchange3}) are the fluxes due to a net
 transfer of mass $\dot{M}$ from one stream to the other. Such 
 transfer must exist at the outer boundaries of a convective zone 
 where the fluid effectively makes a U-turn.

There is no obvious, simple physical prescription for
$\dot{M}$ so we make the assumption that $u=v_{\rm u}=v_{\rm d}$ and
 convective motions are symmetric with respect to the mean
velocity $v$. This approximation is in fact implicit in almost all
convective models to date. The term $\dot{M}$ is then given by equations
(\ref{dotm})--(\ref{massd}) as
\begin{equation}
  \dot{M}=\dpart{\dot{m}}{r}
\mbox{.\\} \end{equation}
  We now have a complete time-dependent description of the
  convective properties of the flow and can investigate
  the characteristics of this model.

\section{Comparison with existing models for  convection}
\label{comparison}

 In this section we consider the approximations made in a number of theories of
 convection. Using the same assumptions, we derive the equations
 for the convective motions in our framework and emphasise the
 characteristics that are peculiar to our formalism.

\subsection{Mixing length theory}
\label{MLT}
  Mixing length theory (MLT) assumes a stationary state for convection
  and makes the quasi-static approximation, $v\simeq0$. In this
  case, all $\Dd{}{t}$ terms can be set equal to zero in the
  difference equations. In addition, we set $\Delta u$, $\Delta
  \epsilon$, $\Delta L$ and $\Delta {\bf R}$ to zero because the processes
  that lead to these terms are usually neglected in MLTs. We also
  neglect the non-local term in the momentum difference equation. The
  difference equations then simplify to
\begin{eqnarray}
\label{MLTdmomentum}
\frac{\Delta \rho}{\rho^2} \dpart{p}{r}&=&2\frac{u^2}{\lambda},\\
\label{MLTdenergy}
\dpart{e}{r}+p\dpart{}{r}\frac1{\rho}&=&-2\frac{\Delta h}{\lambda}
-2\beta\frac{\chi c_p \Delta T}{u\lambda^2}\mbox{ and}\\
\label{MLTdchemistry}
\dpart{{\bf N}}{r}&=&-2\frac{\Delta {\bf N}}{\lambda}.
\end{eqnarray}
  
  Note that equation (\ref{MLTdmomentum}) is not exactly the same as
 in classical MLT~: there is usually a factor of 1/2 multiplying the
 left hand side of this equation to account for the fact that half of
 the work done by the buoyancy force is used to push aside the
 surrounding medium when a convective element rises.

From these equations we extract the convective velocity,
 the temperature and chemical differences between the two
streams (see appendix \ref{dercubic} for a detailed derivation).
These can now be used to express the convective terms in the mean
equations,
\begin{eqnarray}
X_{\rm conv}&=&-\frac1{\rho}\dpart{}{r}(\rho u^2),\\
\dot{m}&=&\frac12 S\rho u,\\
L_{\rm conv}&=&\dot{m}(c_p \Delta T +\bmu'.\Delta {\bf N}),\\
W_{\rm conv}&=&-\frac{u^3}{\lambda}\mbox{ and}\\
{\bf F}_{\rm conv}&=&2\dot{m}\Delta {\bf N}
\end{eqnarray}
  where $\bmu'=T\dth{s}{\bf N}{T,p}+{\bmu}$ and $\bmu$ is the chemical potential (see appendix \ref{dercubic}).

  With these equations, we have derived a MLT that is consistent with
  chemistry. While the excess temperature equation is unchanged, the
  convective velocity now depends on the chemical stratification
  (through the cubic \ref {cubic}) and so does the convective
  luminosity. There is also an additional work term due to the fact
  that we assume a reversible process for the momentum exchange (see
  section \ref{cK86}). We also obtain an explicit change in the
  condition of hydrostatic equilibrium owing to convection.  Finally,
  convection naturally appears as a diffusion process for the
  chemistry.

\subsection{\citet{U67}}
  If we include the D/Dt terms in equations (\ref{MLTdmomentum}) and
  (\ref{MLTdenergy}) and further neglect the chemistry dependence of
  the convective luminosity and velocity we immediately recover the same
  time-dependent version of MLT as \citet{U67} for the excess
  temperature and the convective velocity.

\subsection{\citet{K86}}
\label{cK86}
  \citet{K86} only computes the evolution equation for the convective
  velocity. He uses a diffusion model to compute the
  correlations between velocity perturbations and any other
  perturbation. This model is recovered in our formalism if we
  set $\Dd{}{t}\equiv0$ and $\Delta {\bf R}=\Delta \epsilon=\Delta L=0$ in the
  difference equations for energy and chemistry.

  Furthermore, each term of his equation (25c) for the convective
  kinetic energy corresponds to one term in our equation
  (\ref{dmomentum}), except that his non-local term $\frac1{<\rho>}{\rm
  div}{\bf j}_t$ is different from ours (see paragraph \ref{GNA}), and
  we do not account for his viscous terms (we assumed an inviscid
  fluid).

  Finally, if we compare our equation for the average internal energy
  of the gas, we note that our model misses the heat production owing to
  dissipation of convective motions. This is due to the fact that
  we assume a reversible exchange of momentum so that
  there is no associated heat production. In contrast, \citet{K86}
  assumes viscous dissipation for the convective motions. 

\subsection{\citet{E83}}
\citet{E83} uses his rule of thumb to average the hydrodynamical equations
in order to get evolution equations for the mean fluid and the perturbed
quantities. He then obtains a full set of equations that can be identified
with our mean and difference equations. Our formalism
agrees fairly well with this rule of thumb and hence provides a more
physical basis for it.

  We can recover almost the same equations if we set $\Delta u=0$.
  The local models differ only in the mean energy equation because 
  Eggleton neglects the thermal part of the chemical potentials. However,
  those terms are important when beta decays or electron captures
  occur in degenerate matter.  

  As for the non-local terms, $\dpart{\Delta L}{m}$ in the energy
  difference equation takes the same form of the term in equation (41)
  of \citet{E83} that accounts for the thermal diffusion through the
  front and back of the eddies. However, our non-local term in the
  momentum difference equation does not agree with equation (42) of
  \citet{E83} for the velocity perturbation.

  An interesting point is that \citet{E83} found a term similar to our
  $\Delta {\bf R}$ term.  We have shown that an additional term
  $\Delta \epsilon$ enters into the energy equation. These terms
  account for differential reactivity in the two streams.  However,
  our stationary computations show that these effects are negligible
  as far as the convective Urca process is concerned (see
  section~\ref{discussion}).

 \subsection{\citet{G93}}
\label{GNA}
 
  \citet{G93} use the Boltzmann equation coupled with dynamical equations
to compute a hierarchy of moments for the hydrodynamical equations.
In their framework, our formalism can be recovered
if we specify the distribution
function of the blobs as
\begin{equation}
f_A(t, z, v, T)=\frac1{S} 
[S_{\rm u}\rho_{\rm u} \delta(v+v_{\rm u},T_{\rm u})
+S_{\rm d}\rho_{\rm d} \delta(v-v_{\rm d},T_{\rm d})]
\mbox{. } \end{equation}
This allows a direct comparison between the two formalisms.
With this definition, our variables $v$, $\rho$ and $T$ correspond to their
variables $\bar{v}$, $\bar{\rho}$, and $\bar{T}$.

The zeroth-order equations of their hierarchy can be directly compared
with our mean equations.  The velocity equation is found to differ by
terms negligible in the subsonic regime. Their temperature equation
without chemistry has an additional source term due to viscous
dissipation just as \citet{K86}. Their temperature equation with
chemistry misses correction terms in the luminosity, the work term and
the source term. This is due to their use of a dynamical equation for
the entropy which does not account for the changes owing to chemical
evolution. This suggests that energy conservation may not hold in
their case when chemistry is included.

Their higher-order equations can also be compared with our difference
equations, although this is less straightforward. We did such a comparison
but only for the velocity difference equation. We then obtain the same
non-local term as \citet{K86}: their $\frac1{\bar{\rho}}
\dpart{}{z}(\bar{\rho}\bar{w^3})$ corresponds to $\frac1{<\rho>}{\rm
  div}{\bf j}_t$ of \citet{K86}. If we substitute our distribution
function $f_A$ in this term, we obtain
\begin{equation}
\frac2{\rho}\dpart{}{z}(\rho u^2 \Delta u)
\end{equation}
which should be compared to our
\begin{equation}
2u^2\dpart{\Delta u}{r}+u^3\dpart{}{r}\frac{\Delta \rho}{\rho}
\end{equation}
when $v=0$. 
Our term contains $\Delta \rho$ and theirs contains the spatial
derivative of the convective velocity.
On the other hand, our treatments agree as far 
the chemical dependence of the convective velocity is concerned.

\section{The convective Urca process}
In this section we first present the basic nuclear reactions
responsible for the Urca process. Because previous studies have mainly
concentrated on the entropy equation we derive it  in our
formalism and then examine the influence of the Urca nuclei on
convection.

\subsection{Urca reactions}
  Urca reactions involve pairs of nuclei of the form
  $(_A^{Z+1}$M,$_A^Z$D$)$ where A is an odd number. The member of a pair
  with an additional proton is called the mother (M), while the other
  one is referred to as the daughter (D). Electron capture and beta decay  
  turn one into the other~:
\begin{equation}
\mbox{Electron capture~:    M}  +e^- \rightarrow \mbox{D}+\nu
\end{equation}
\begin{equation}
\mbox{Beta decay~:          D}    \rightarrow \mbox{M}+e^-+\bar{\nu}
\end{equation}

  \citet{TC70} give the reaction rates for electron capture and beta
  decay $\lambda^+$ and $\lambda^-$ per nucleus. They
  also provide the corresponding neutrino losses 
  $L^+$ and $L^-$ per nucleus.  The typical time-scale for Urca
  reactions is 10$^5$~s.
 These reaction rates depend mainly on the chemical potential $\mu_e$ of
 the electrons. This is mainly on the mass density $\rho$ of the
 degenerate matter. Each Urca pair has a threshold energy $\mu_{\rm
 th}$ above which significant electron captures can occur.  We
 approximate the Coulomb corrections to these threshold energies in
 the same way as \citet{G96}.  When $\mu_e>\mu_{\rm th}$ (i.e.
 $\rho>\rho_{\rm th}$) electron captures quickly turn the Urca matter
 into daughter nuclei.  When $\mu_e<\mu_{\rm th}$ (i.e. $\rho<\rho_{\rm
 th}$) beta decays quickly turn it into mother nuclei. When
 $\mu_e\simeq \mu_{\rm th}$ ($\rho\simeq\rho_{\rm th}$) both
 reactions are significant and the Urca matter quickly evolves into a
 mixture of mother and daughter nuclei. A shell on which
 $\rho=\rho_{\rm th}$ is called an Urca shell.

 When the Urca matter is in chemical equilibrium both reactions take
 place at the same rate, which is highest near the location of the
 Urca shell.  Because both reactions emit neutrinos, this leads to neutrino
 cooling which is strongest at the Urca shell. When the
 Urca matter is far from equilibrium, nuclear heating takes place at the
 same time and usually dominates over the neutrino cooling
 \citep[see][ and section \ref{netheating} below]{B73}.

 The pair $^{23}{\rm Na}/^{23}{\rm Ne}$ is a typical example 
 and perhaps the most active Urca pair in massive white
 dwarfs. Its threshold energy is $\mu_{\rm th}=4.38$~MeV, which
 corresponds to $\rho_{\rm th}=1.7\times 10^9$~g.cm$^{-3}$ in purely
 degenerate matter.  This density is slightly below the density for
 carbon ignition in a C+O white dwarf accreting at a rate of
 $10^{-7}~$M$_{\sun}$yr$^{-1}$. As a consequence, the growing convective
 core soon engulfs the corresponding Urca shell after carbon has
 ignited in the centre.

  Here we assume that the stellar matter contains a fixed
  number $N_{\rm U}$ of Urca nuclei (per unit mass) of a given
  pair. We define $N_{\rm M}$ and $N_{\rm D}$ as the corresponding
  number of mother and daughter nuclei so that $N_{\rm U}=N_{\rm
  M}+N_{\rm D}$.  Then the rate of change of mother and daughter
  nuclei per unit mass becomes
\begin{equation}
R_{\rm M}=-R_{\rm D}=-\lambda^+ N_{\rm M} + \lambda^- N_{\rm D}.
\end{equation}
  If we write $N^*_{\rm M}=N_{\rm U} \lambda^-/(\lambda^+ +
  \lambda^-)=N_{\rm U}-N_{\rm D}^*$ as the number of mother nuclei per
  unit mass at chemical equilibrium we can rewrite the reaction rate
  as
\begin{equation}
\label{RM}
R_{\rm M}=-(\lambda^++\lambda^-)(N_{\rm M}-N_{\rm M}^*).
\end{equation}

  Useful explicit expressions for the various 
  thermodynamical properties of the Urca nuclei are
\begin{eqnarray}
\mu_{\rm M}&=&(Z+1)\mu_e+{k}T (a+\ln N_{\rm M}),\\
\mu_{\rm D}&=&Z\mu_e+{k}T (a+\ln N_{\rm D}),\\
\mu'_{\rm M}&\simeq&(Z+1)\mu_e+\frac12{k}T,\\
\mu'_{\rm D}&\simeq&Z\mu_e+\frac12{k}T,\\
\mu''_{\rm M}&=&-(Z+1)N_{\rm M}/N_e \mbox{and}\\
\mu''_{\rm D}&=&-ZN_{\rm D}/N_e,
\end{eqnarray}
  where $N_e$ is the total number of electrons per unit mass, $k$ is the
  Boltzmann constant and $a$ is a combination of temperature and
  density logarithms \citep[see][]{P95}. The approximations for
  $\mu'_{\rm M}$ and $\mu'_{\rm D}$ are obtained by neglect of the
  electron and radiation contributions to the entropy.  In the
  strongly degenerate case the $kT$ terms are usually negligible.

\subsection{Convective velocity}
\label{Urcavel}

  In appendix \ref{stability} (equation \ref{uconv}) we derive the
  approximation which relates the convective velocity to the
  temperature and chemical gradients
\begin{equation}
u=u_1\sqrt{\delta (\nabla-\nabla_a)-\bmu''.{\bf \bnabla_N}}
\mbox{,\\} \end{equation}
  where the term depending on the Urca process is
\begin{equation}
\label{uchem1}
\bmu''.{\bf \bnabla_N}=-\frac1{N_e}[(Z+1)N_{\rm M} \nabla_{\rm
M}+ZN_{\rm D} \nabla_{\rm D}] \mbox{.\\} \end{equation} Since $N_{\rm
U}$ is assumed to be uniform, $N_{\rm M} \nabla_{\rm M}+N_{\rm D}
\nabla_{\rm D}=0$.  Hence relation (\ref{uchem1}) becomes
\begin{equation}
\label{uchem2}
\bmu''.{\bf \bnabla_N}=-\frac{N_{\rm M}}{N_e}\nabla_{\rm M}
\mbox{.\\} \end{equation}
Electron captures are much stronger in the centre of the star
than in the outer regions so the mother fraction generally increases
outwards, $\bmu''.{\bf \bnabla_N}>0$ and the effect of the
presence of Urca pairs is to {\it reduce} the convective velocity. In
sections \ref{centre} and \ref{stationary} we show that this effect
can actually be quite strong and may even inhibit convection. Note
that, in the case where $N_{\rm U}$ is not uniform, the sign of the 
$\bmu''.{\bf \bnabla_N}$ term may change.

\subsection{Entropy equation}
\label{entropy}

  One of the biggest uncertainties in previous models for the convective
  Urca process was the form of the equation for the evolution of the
  entropy $s$ in the presence of convection. Here we derive the entropy
  equation in the one-flow limit. The small $\Delta$ approximation
  allows us to use thermodynamic relations for the mean fluid and we
  can transform the left-hand side of equation (\ref{smenergy}) so that
\begin{equation}
T\Dd{s}{t}+\bmu\Dd{{\bf N}}{t}=\epsilon-\dpart{L}{m}
-\dpart{L_{\rm conv}}{m}+W_{\rm conv}.
\end{equation}
  We then use equation (\ref{smchemistry}) and the relation $\Delta
  h-\bmu.\Delta{\bf N}=T\Delta s$ (because $\Delta p=0$) to write the equation
for the mean entropy as
\begin{equation}
\label{smentropy}
T\Dd{s}{t}=\epsilon'-\dpart{L}{m}
-\dpart{L'_{\rm conv}}{m}+W'_{\rm conv},
\end{equation}
 where we define
\begin{equation}
\epsilon'=\epsilon-\bmu.{\bf R}
\mbox{,} \end{equation}
\begin{equation}
L'_{\rm conv}=2\dot{m}T\Delta s
 \end{equation}
and
\begin{equation}
W'_{\rm conv}=W_{\rm conv}-2\dot{m}\Delta{\bf N}.\dpart{\bmu}{m}
\mbox{.\\} \end{equation}
  The entropy equation then takes a form that is similar to the energy
  equation but with different definitions for the net heating, the
  convective luminosity and the work. We next consider how these terms
  are affected by the chemical state of the Urca matter.

\subsubsection{Net heat generation}

\label{netheating}
\begin{figure}
\centerline{
\psfig{file=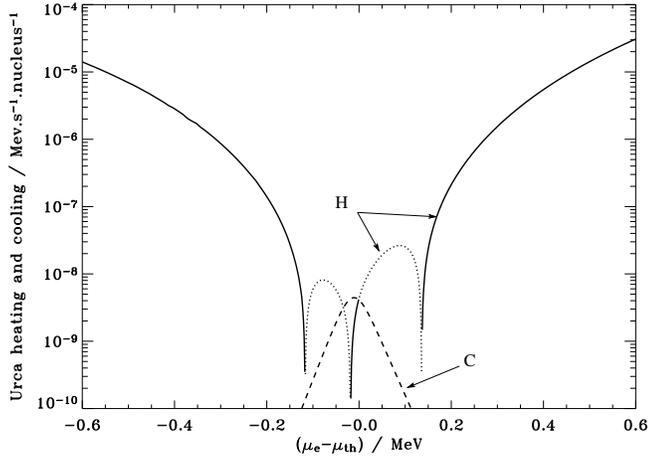,angle=-90,width=9cm}
} \caption{Functions $|H|$ (solid and dotted lines) and $-C$ (dashed line)
for the $^{23}{\rm Na}/^{23}{\rm Ne}$ Urca pair at a temperature of
$3\times 10^8$~K. The solid portions of $|H|$ indicate where 
$H$ and $N_{\rm M}-N^*_{\rm M}$ have the same sign (heating)
and dotted otherwise (cooling).}
\label{HC}
\end{figure}

The net heating due to Urca reactions is
\begin{equation}
\epsilon_{\rm U}'=\mu_{\rm th}R_{\rm M}-L^+N_{\rm M}-L^-N_{\rm D}-\mu_{\rm M} R_{\rm M}-\mu_{\rm D} R_{\rm D}\mbox{.}
\end{equation}
In the very degenerate case, $\mu_{\rm M}\simeq(Z+1)\mu_e$ and $\mu_{\rm D}\simeq
Z\mu_e$, we obtain
\begin{equation}
\epsilon_{\rm U}'=N_{\rm U}C+(N_{\rm M}-N_{\rm M}^*)H,
\end{equation}
where 
\begin{eqnarray}
C&=&-\frac{L^+\lambda^-+L^-\lambda^+}{\lambda^++\lambda^-}\\
H&=&-L^++L^-+(\mu_e-\mu_{\rm th})(\lambda^++\lambda^-)
\mbox{.}\end{eqnarray} In figure \ref{HC} we plot $|C|$ and $|H|$ for
the Urca pair $^{23}{\rm Na}/^{23}{\rm Ne}$.  The net Urca heating
$\epsilon_{\rm U}'$ is the sum of two terms. At chemical equilibrium,
only the first term remains. This is always negative and so causes
cooling.  The sign of the second term depends on the signs of $N_{\rm
M}-N_{\rm M}^*$ and $H$. The term $N_{\rm M}-N_{\rm M}^*$ is likely to
be positive below the Urca shell and negative above it.  The value of
$H$ has the same sign except close to the Urca shell. Hence, the second term
$(N_{\rm M}-N_{\rm M}^*)H$ is positive, implying heating, except close
to the Urca shell.

The relative magnitude of the second to the first term is proportional
to the departure from chemical equilibrium $(N_{\rm M}-N_{\rm
M}^*)/N_{\rm U}$. One can see from figure (\ref{HC}) that the heating
can be balanced by neutrino losses only if the system is close to
chemical equilibrium because, generally, $|H|\gg|C|$.  In a convective
region mixing puts the Urca abundances slightly (or strongly
if convection is very efficient) outside equilibrium. The amount of
convective mixing is therefore crucial for computing the net Urca
heating. Generally, the net effect is cooling close to the Urca shell
and heating far away from it.

\subsubsection{Convective luminosity}

 The contribution to $L'_{\rm conv}$ associated with the Urca pairs is
\begin{equation}
L'_{\rm U}=2\dot{m}(\bmu-\bmu').\Delta {\bf N}= 2\dot{m}{\rm
k}T\ln(N_{\rm D}/N_{\rm M})\Delta N_{\rm M} \mbox{.\\} \end{equation}
The quantity one needs to compare with $\epsilon_{\rm U}'$ and
$W'_{\rm conv}$ is in fact the mass derivative of the luminosity
$\dpart{L'_{\rm U}}{m}$.  This term is usually negligible in the very
degenerate case. However, at convective boundaries, the derivative
$\dpart{\dot{m}}{m}$ can be large and make this term
relatively more important.

\subsubsection{Work term}
Here we compute the contribution of the Urca pairs to the work term in
the entropy equation,
\begin{equation}
W'_{\rm U}=2\dot{m}\Delta{\bf N}.\dpart{\bmu}{m}=2\dot{m}\Delta N_{\rm M} \dpart{\mu_e}{m}\mbox{.}
\end{equation}
Because the Urca matter is richer in mother nuclei in the outer parts of
the star, $\Delta N_{\rm M}>0$, and because the density is decreasing outwards,
$\mu_e$ is decreasing outwards. Hence, $W'_{\rm U}$ is always {\it
negative}.\\ 
We note that $W'_{\rm U}=F_{\rm M}\dpart{\mu_e}{m}$ is identical to the work
term \citet{I78b} designed for his computations although he uses
it in the energy equation and does not specify what convective
velocity or luminosity he adopts.\\
Finally, no heat production is associated with our chemical exchange
model. Indeed, we assume that a reversible process is responsible for
the mixing of chemical species between both streams. An irreversible
process for this chemical mixing could give rise to a heating term
which could balance part (or all) of this additional work term.

\subsection{Criterion for convection at the centre of the star}
\label{centre}
Let us now assume that we know the composition, temperature and density at
the centre of a star. We can then derive a criterion for whether
there is stationary convection and deduce an upper bound to the amount
of mixing at the centre of a convective Urca core.

  We consider a very small sphere of mass $m$ at the centre of the
  star.  In a stationary state the total luminosity at the edge of
  this sphere must balance the energy production inside the sphere. If
  we assume that all the energy is carried out by convection we can
  write the convective luminosity
\begin{equation}
  \epsilon m=L_{\rm conv}=2\Delta h \dot{m}
=2(c_pT\Delta \ln T+\bmu'.\Delta{\bf N})\dot{m}
\mbox{.}
\end{equation}
  We obtain a similar expression for the net number of particles
of each kind flowing away from this sphere 
\begin{equation}
{\bf R}m={\bf F}_{\rm conv}=2\Delta {\bf N}\dot{m}.
\end{equation}

Combining these two equations we obtain the density difference
at the edge of the sphere
\begin{equation}
 \Delta \ln \rho=\frac{m}{2\dot{m}}
\left(-\delta \frac{\epsilon-\bmu'.{\bf R}}{c_{\rm P} T}+
\sum_j\frac{mu_j''R_j}{N_j}
\right)
.\end{equation}

\begin{figure}
\centerline{
\psfig{file=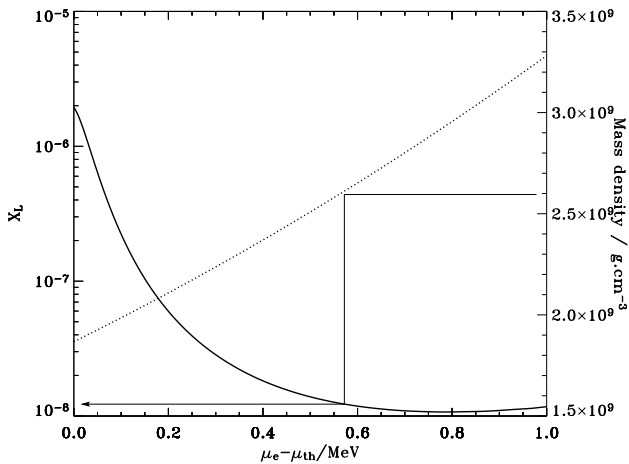,angle=-90,width=9cm}
} 
 \caption{$X_{\rm L}$ (solid line) and $\rho$ (dotted line) against
$\mu_e-\mu_{\rm th}$ for the $^{23}{\rm Na}/^{23}{\rm Ne}$ Urca pair at a
temperature of $3\times 10^8$~K.}
\label{XL}
\end{figure}

A stationary convective state exists at the centre if and only if
$\Delta \rho<0$. This translates into an upper limit for the mother
fraction at the centre of an Urca convective core. Using the relations
$\mu_j'\simeq Z_j\mu_e$ and $\mu_j''=-Z_jN_j/N_e$, where $N_e$ is the 
number of electrons per unit mass, we obtain
\begin{equation}
\label{inequ}
N_{\rm M}-N_{\rm M}^*<\delta\frac{\epsilon^*}{c_{\rm P}T}
\left(\frac{\lambda^++\lambda^-}{N_e}-\frac{\delta H}{c_{\rm P}T}\right)^{-1}
=N_{\rm L}=AX_{\rm L}
\mbox{,}
\end{equation}
where $\epsilon^*$ is the net heating at Urca chemical equilibrium
(equal to the heating from carbon burning at the centre) and
$N_{\rm L}$ and $X_{\rm L}$ are defined by equation (\ref{inequ}). 
 $X_{\rm L}$
is the minimum mass fraction of Urca pairs which has a
significant effect on convection. It generally has a fairly small value
(see figure \ref{XL}).

When $N_{\rm U}\gg N_{\rm L}$ the Urca pairs must be close to chemical
equilibrium. In other words, the convective core can be only very
slightly mixed.  But if they were in chemical equilibrium carbon
burning alone would produce a large buoyancy and hence drive strong
mixing. Therefore, the Urca composition has to adjust itself to balance the
heat from carbon burning and produce almost zero convective
velocities.  Hence, the inequality (\ref{inequ}) is nearly an equality
in practice and $N_{\rm M}-N_{\rm M}^*\simeq N_{\rm L}$, a result
that is verified in our simulations of stationary 
convective cores (see Section~\ref{results}).

\section{Stationary convective Urca cores}
\label{stationary}
If we set all $\dpart{}{t}$ derivatives equal to zero in equations
(\ref{massu})--(\ref{chemistryd}), we obtain a system of coupled
ordinary differential equations.  If we specify the state variables at
the centre of the star, we can integrate these equations outward,
using a shooting method, and calculate the hydrostatic profile of
the star.

To obtain a guess for the state variables at the centre, we run a very
simple time-dependent model of an accreting white dwarf without Urca
nuclei. We then compute the stationary state of stationary convective
cores for different Urca compositions and different versions of our
convective model.

\subsection{Time-dependent model}
\label{tdep}

We used the \citet{E71} stellar evolution code to calculate the
evolution of a white dwarf, composed entirely of $^{12}$C, $^{16}$O
and $^{20}$Ne (with mass fractions $^{12}{\rm C}=0.25$, $^{16}{\rm O}=
0.73$, $^{20}{\rm Ne}=0.02$) accreting matter at a rate of
$10^{-7} M_{\sun}$/yr. The initial mass was taken as 
1~$M_{\sun}$. Only the carbon-burning
reaction $^{12}$C($^{12}$C,$\alpha$)$^{20}$Ne immediately followed
by $^{12}$C($\alpha$,$\gamma)^{16}$O was
taken into account. The convective model used in the Eggleton code is
standard MLT and we use the approximate equation of state of
\citet{P95} assuming complete ionisation.

\begin{figure}
\centerline{
\psfig{file=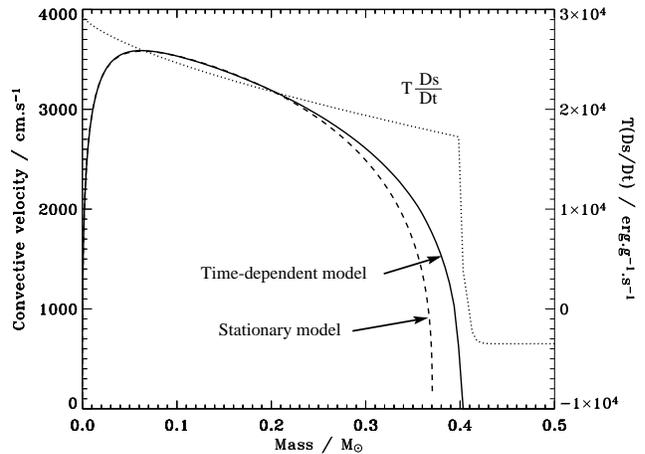,angle=-90,width=9cm}
}  

 \caption{Convective velocity profile through the core. The solid line is
 for the time-dependent simulation and the dashed line is for the hydrostatic
 simulation. We also plot $T\Dd{s}{t}$ in the time-dependent simulation
 (dotted line). }
\label{cct}
\end{figure}

  We stop the computation during the carbon flash when the convective
  core has reached a mass of $0.4~M_{\sun}$. We plot the convective
  velocity of this core against mass in figure (\ref{cct}). At this
  point, the threshold density for the $^{23}{\rm
  Na}/^{23}{\rm Ne}$ Urca pair is in the middle of the convective
  core. The central density is $2.6\times 10^9$g.cm$^{-3}$ and the central
  temperature is $3.1\times 10^8$~K.

If we were to integrate the equations of hydrostatic equilibrium from this
central state using MLT, we would obtain a fully convective star. The
reason is that the core is actually being heated and that the term $T
\Dd{s}{t}$ is non-zero. This term is negligible in the carbon-burning
region but is significant in the outer part of the convective
region. However, the effect of convection is to homogenise the entropy
profile and so $T \Dd{s}{t}$ is rather uniform in the
convective region. We can therefore use its value to offset the
nuclear heating when we compute the hydrostatic profile.  This brings
the stationary convective profile very close to the time-dependent one
(see figure \ref{cct}).

\subsection{The shooting method}
 Setting all the $\dpart{}{t}$ terms to zero in the equations for the
  two streams, we obtain a system of coupled ordinary differential
  equations, which we integrate numerically from the centre of the
  star. 
As central boundary conditions for temperature, density, C, O and
$^{20}$Ne mass fractions we take the results of the time-dependent model.
For each value of $N_{\rm U}$, we make an initial guess for 
the number density $N_{\rm M}$. The estimate for $N_{\rm M}$ is then
iteratively improved by successive  outward integrations until the condition
$\Delta N_{\rm M}=0$ is satisfied at the outer edge of the convective region.

\subsection{Results}
\label{results}
  We compute stationary convective cores for different mass fractions
  of Urca pairs $X_{\rm U}=23 \times N_{\rm U}=0$ , $10^{-12}$,
  $10^{-9}$, $10^{-8}$, $10^{-6}$, $10^{-3}$. A value of $X_{\rm
  U}=10^{-3}$ would require a very efficient conversion of $^{20}$Ne
  into Urca pairs and hence gives a reasonable upper limit for the
  possible abundance of Urca pairs.
 
\subsubsection{Without Urca nuclei}
\label{withouturca}

\begin{figure}
\centerline{
\psfig{file=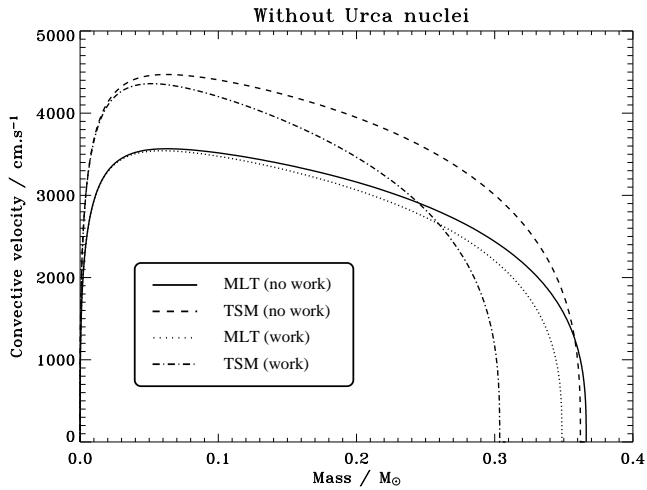,angle=-90,width=9cm}
}   

 \caption{Convective velocity profile of different models of
 stationary cores {\it without} Urca nuclei.  Solid and dashed lines
 are for MLT and TSM models with viscous dissipation of
 momentum. Dotted and dash-dotted lines for MLT and TSM models with the
 work term $W_{\rm conv}$.  }

\label{noUrca}
\end{figure}

  For a pure C+O+$^{20}$Ne mixture, we compare the velocity profiles
  of the stationary convective core given by MLT and the two-stream
  model (TSM). In the TSM, we first set the work term $W_{\rm conv}$
  equal to 0.  Figure (\ref{noUrca}) shows that the velocities differ
  by a factor of $\sqrt{2}$. This is the only difference between the
  two models and it can be traced back to the factor 2 in
  equation~(\ref{MLTdmomentum}). Indeed, MLT has a factor 4 instead
  because it assumes that half of the work done by the buoyancy forces
  is used to push aside the surrounding medium when a convective
  element rises.

  When we put the work term back in MLT or TSM the resulting
  convective core shrinks. Less convection is needed to carry out the
  C burning energy. This is not surprising because adding the work term
  is equivalent to suppressing the viscous heat produced by the
  dissipation of the drift motions.

\subsubsection{Very low Urca abundance}
\begin{figure}
\centerline{
\psfig{file=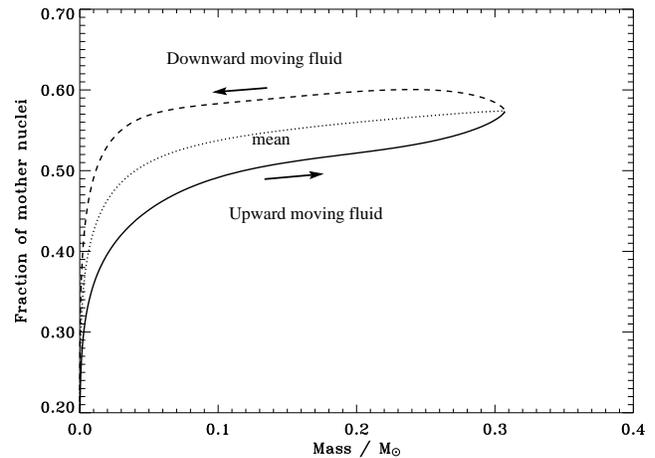,angle=-90,width=9cm}
}   

 \caption{Fraction of mother nuclei $X_{\rm M}/X_{\rm U}$ against mass in the upward and
downward moving streams for $X_{\rm U}<10^{-9}$. The mean is also indicated. 
The model is the two-stream model (TSM) including the work term.}
\label{lowmix}
\end{figure}

  For $X_{\rm U}<10^{-9}$, the Urca nuclei do not have an impact on the convective
  velocity. They are mixed passively through the convective region.
  Figure \ref{lowmix} shows the relative abundance of mother nuclei
  $N_{\rm M}/N_{\rm U}$ in both streams. Mother nuclei come from above the Urca
  shell. They capture electrons as they descend below the Urca
  shell and are converted into daughter nuclei. As these rise back 
  above the Urca shell, they emit electrons and the number of 
  daughter nuclei rises again. Finally, they  cycle back down through the Urca shell.

  The maximum relative difference of composition between both fluids
  is 17\%. This suggests that the one-stream approximation may well
  be adequate.

\subsubsection{High Urca abundance}
\begin{figure}
\centerline{
\psfig{file=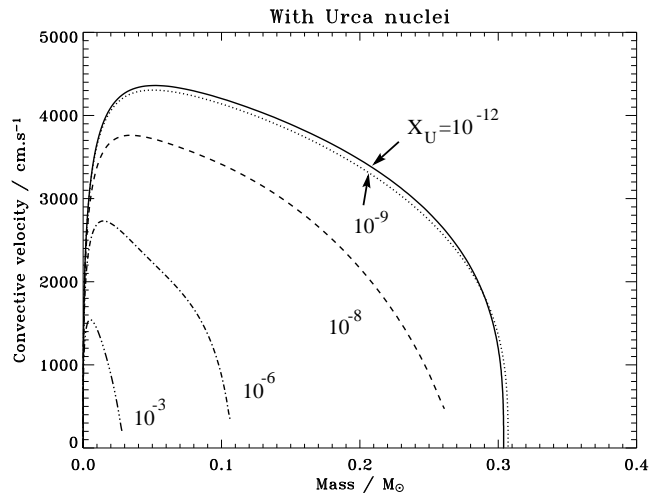,angle=-90,width=9cm}
}   

 \caption{Convective velocity profile of different models of
 stationary cores {\it with} Urca nuclei.  The different values for
 $X_{\rm U}$ (the total mass fraction of Urca nuclei) are indicated next to
 each curve.  }
\label{HighUrca}
\end{figure}

  For $X_{\rm U}>10^{-8}\simeq X_{\rm L}$ our shooting method is not
  able to find a stationary state.  When the mother mass fraction at
  the centre is too high the relative composition difference diverges
  as the convective velocity tends to zero. When it is too low the
  convective velocity reaches a minimum and then rises again up to 
  the edge of the white dwarf. Figure \ref{HighUrca} shows the
  convective velocity profiles in the latter case and demonstrates how
  drastically the Urca nuclei can affect the velocity profile.

\begin{figure}
\centerline{
\psfig{file=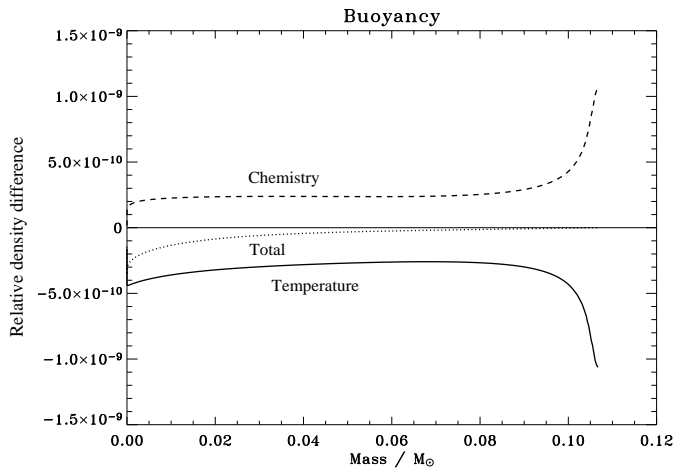,angle=-90,width=9cm}
}  

 \caption{Density difference $\Delta \ln \rho$ through the convective
 core for a low central mother fraction case.  $\Delta \ln
 \rho=-\delta \Delta \ln T+ \bmu''.\Delta \ln{\bf N}$. It is the
 sum of a temperature-dependent term and a chemistry-dependent
 term. We plot $-\delta \Delta \ln T$ (solid line), $\bmu''.\Delta
 \ln{\bf N}$ (dashed line) and $\Delta \ln \rho$ (dotted line).}
\label{drho}
\end{figure}

 We illustrate this effect by plotting the density difference $\Delta
 \ln \rho$ in figure \ref{drho} for a case where $X_{\rm M}$ at the
 centre is too low. Indeed, the density difference controls the
 buoyancy force and hence the convective velocity.  This figure shows
 the competing dependence of $\Delta \ln \rho$ on the temperature and
 on the chemistry (the Urca nuclei). As was previously noted in
 section~\ref{Urcavel}, the chemical part of the density difference is
 positive and stabilises convective motions. In the outer parts of the
 convective region the temperature and chemical dependence cancel
 each other to yield a very small convective velocity. This suggests
 that the criterion for semi-convection might be fulfilled in the
 region just above the point where $u$ vanishes.

\begin{figure}
\centerline{
\psfig{file=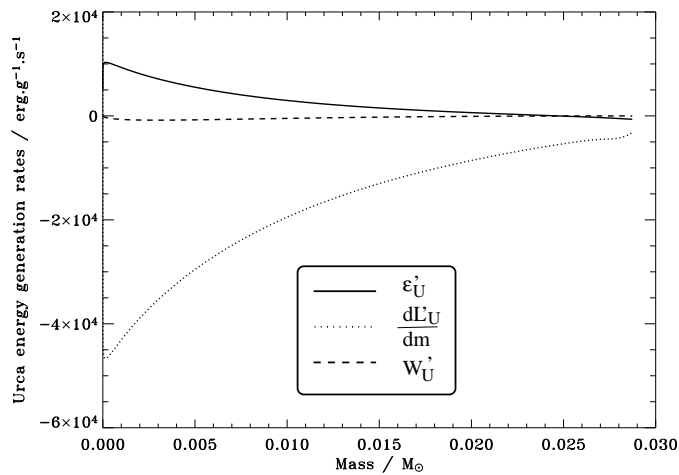,angle=-90,width=9cm}
} 
\caption{ Quantities $\epsilon'_{\rm U}$ (solid), $\dpart{L'_{\rm U}}{m}$ (dotted) and
$W'_{\rm U}$ (dashed) are plotted for a low central mother fraction
case of model $X_{\rm U}=10^{-3}$ (TSM with work term). The total net
energy generation rate is greater than $5\times
10^4$~erg.g$^{-1}$.s$^{-1}$. It is dominated by the energy generation
due to carbon burning.}
\label{sconv}
\end{figure}

In degenerate matter the mass density is tied to the electron
abundance. Hence, all reactions that change the number of electrons have
an effect on the buoyancy.  The terms $\epsilon_{\rm U}$,
$\dpart{L'_{\rm U}}{m}$ and $W'_{\rm U}$ seem to be of secondary
importance compared to the change in the convective velocity. We plot
them in figure \ref{sconv}. They are relatively unimportant
compared to the energy generation rate due to carbon burning.

\section{Discussion}
\label{discussion}

  In principle the formalism we have derived is sufficiently general
  that it can be used with any model for the exchange (or diffusion)
  between the two streams.  It automatically
  guarantees conservation of the chemical species and energy and
  allows time dependence.  The formalism also accounts for the
  interactions between the global contraction or expansion (with
  velocity $v$) of the star and the convective (or drifting) velocity
  $u$.  In fact, the drifting velocities for the upward and downward
  motion ($v_{\rm u}$ and $v_{\rm d}$) do not even have to be equal.
  
  But it is only a formalism and requires a model for all horizontal
  motions/exchanges. In section \ref{model} we suggested a model that
  is easy to implement and which can be used to compare our new
  formalism with previously derived theories of convection. However,
  this has a few limitations.  

First, we used a very simple model {(where $\Delta u$ was set equal to
zero)} for the net mass transfer $\dot{M}$ between the upward and
downward moving fluids.  This eliminates the main non-local terms
involving $\Delta u$ in equation (\ref{dmomentum}) and hence disables
the effects of convective overshooting. To account for overshooting in
a self-consistent way requires an {\it a priori} physical model for the
exchange term $\dot{M}$.  However, one can investigate the effects of
overshooting within our framework by using a necessarily somewhat {\it
ad hoc} prescription for the convective velocity which allows a finite
convective speed beyond the formally convective region according to
the Ledoux criterion. 

Second, we chose reversible processes for the exchange of momentum and
the mixing of chemical species. Irreversible processes would dissipate
part (if not all) of the work done as heat.  The cooling terms $W_{\rm
conv}$ and $W'_{\rm U}$ would then have lower (possibly zero) values.
For example, the Reynolds numbers in convective regions are so large
that there is almost certainly a large amount of turbulence.  To
account properly for turbulent dissipation in the derivation of our
model would require the inclusion of a finite viscosity introduced by
the turbulent cascade. This is not easily done. However, one can
investigate the possible outcome of dissipation by switching the work
terms on and off. In section \ref{withouturca} (Figure \ref{noUrca}),
we investigated the effect of setting $W_{\rm conv}=0$ and found
differences of up to 20\%.  Furthermore, setting $W'_{\rm U}=0$ would
most likely only have a small effect because $W'_{\rm U}$ is already
small.
 
\subsection{Stationary cores}

  As a first illustration of our two-stream formalism we computed
  {\it stationary} convective cores. However, the convective core
  during the carbon flash is growing very rapidly. Indeed, the time
  dependence does matter at least in the energy equation, as was shown
  in section \ref{tdep}.  Our stationary cores may therefore only be
  very rough approximations of growing convective cores. Moreover, in
  the present study we used arbitrary abundances for the Urca
  nuclei. Most of the Urca nuclei are by-products of carbon
  burning. To compute their abundances in a self-consistent manner
  involves quite an extended network, as was shown by \citet{I78a}.

  Despite these limitations, the results already help to shed some
  light on the convective Urca process and prove useful in calibrating
  numerical aspects in the implementation of the method.  For example,
  the terms $\Delta {\bf R}$, $\Delta \epsilon$, and $\Delta L$, as
  well as the non-local term in equation (\ref{dmomentum}), are found
  to be second order terms in our computations of stationary Urca
  cores.  Furthermore,
  the chemical composition differences between the two streams are
  generally small, except possibly at the outer edge of a convective
  core with a high Urca abundance.  This may
  provide some justification for the use of the one-stream approximation to
  describe the convective Urca process in a stellar evolution code.

\subsection{Future work}

  The next step is to implement the two-stream model (or a more
  simple, but appropriate approximation) in a stellar evolution
  code. Coupled with a suitable nuclear reaction network, this will
  allow us to follow the evolution of the core in a fully
  self-consistent manner to the runaway phase in a SN Ia and to
  determine the physical and chemical conditions in the core at the
  time of the explosion. This will provide the ignition conditions,
  the thermodynamic properties and the location of the ignition point
  for explosion calculations \citep[e.g.][]{H04}. Moreover, with this
  tool, we shall be able to systematically address the dependence of
  the ignition conditions on the overall metallicity, the initial C/O
  ratio, the white dwarf accretion rate, the initial mass of the white
  dwarf etc.  

  In this context, we note the importance of the neutron excess at the
  time of the explosion. Timmes, Brown \& Truran (2003) have recently
  emphasised the role of the initial $^{22}$Ne abundance, which they
  argued was determined by the metallicity, introducing a metallicity
  dependence for SNe Ia. However, the neutron excess itself is
  affected by the electron captures and emissions in the simmering
  phase preceding the nuclear runaway.  If we take as indicative the
  Urca abundances derived by \citet{I82} (table 4) at the end of his
  computations, we can compute the number of additional neutrons
  introduced by the Urca reactions alone on Ne isotopes,
  $n_U=3Y(^{23}$Ne$)+5Y(^{25}$Ne$)=3.6\times10^{-3}$.  The
  corresponding number of additional neutrons caused by $^{22}$Ne
  resulting from solar abundances in the white dwarf progenitor is
  $n_\odot=2Y_\odot(^{22}$Ne$)=2.5\times10^{-3}$.  Hence the effect of
  the Urca isotopes of Ne on the neutron excess can be even larger
  than that of fossil $^{22}$Ne. This provides another illustration
  for the importance of a proper treatment of the Urca process for
  answering some of the fundamental, unsolved questions concerning SNe
  Ia.

\section{Summary and conclusions}

  We have derived a two-stream formalism which carefully addresses
  the energy and chemical budgets in a convective region. In addition,
  it allows time dependence and describes the interaction of
  convection with the general motions of the star. We illustrated this
  formalism with a simple model and compared the resulting theory of
  convection to existing theories.  We also derived a one-stream limit
  approximation which will be easy to implement in a stellar evolution
  code as an extension of classical MLT.

  We then applied this formalism to the convective Urca process and
  derived the entropy equation which has been central to previous
  discussions of the Urca process, and computed the convective
  velocity. We showed that the net heating effect of the Urca process
  strongly depends on the state of mixing of the convective core, for
  which we provide an estimate. Urca reactions generally tend to
  reduce the effects of buoyancy. More generally we show that, in
  degenerate matter, reactions that change the number of electrons
  have a direct influence on the convective velocity.

  As an illustration of our model we computed {\it stationary}
  convective cores. These computations show that, even for a very small
  Urca fraction, convective velocities are strongly modified compared
  to the case without Urca nuclei. They also show that convective Urca
  cores are unlikely to be in a stationary state. Hence,
  time-dependent computations with a full nuclear reaction network are
  needed to provide the final answer to the question, ``what is the
  influence of the convective Urca process on the ignition conditions
  in type Ia supernovae?''

\appendix
\section{Stability analysis of convection}
\label{stability}

In this appendix we analyse the stationary states of convection when the
evolution of the average quantities is very slow. We first investigate
the possible available stationary states and then study their
linear stability.

In the following we assume that $\Delta {\bf R}=\Delta \epsilon=\Delta
L=\Delta u=0$ and neglect $v$ and non-local terms in the difference
equations.

\subsection{The stationary states} 
The radiative state ($u=0$) is always a stationary solution of our
equations.  The other stationary convective states are solutions
of the system of equations (\ref{MLTdmomentum})--(\ref{MLTdchemistry}) when
$u=0$ has already been factored out. We now compute the solutions
of this system and examine their existence and possible multiplicity.

\subsubsection{The cubic equation for the convective velocity}
\label{dercubic}

Standard thermodynamics gives
\begin{displaymath}
\dpart{e}{r}+p\dpart{}{r}\frac1{\rho}
=T\dpart{s}{r}+{\bmu}.\dpart{{\bf N}}{r}
\end{displaymath}
\begin{displaymath}
=Tc_p(\dpart{\ln T}{r}-\nabla_a\dpart{\ln{p}}{r})
+{\bmu}'.\dpart{{\bf N}}{r}=
\end{displaymath}
\begin{equation}
-\frac{T c_{\rm P}}{H_p}(\nabla-\nabla_a)+{\bmu}'.\dpart{\bf N}{r},
\end{equation}
 where $s$ is the specific entropy, $\nabla_a=\dth{\ln T}{\ln
 P}{s,{\bf N}}$, $c_p=\dth{h}{T}{p,{\bf N}}$, $\bmu$ are the chemical
potentials and $\bmu'={\bmu}+T\dth{s}{\bf N}{T,p}$. We also have
\begin{equation}
\Delta h=c_p \Delta T+{\bmu}'.\Delta {\bf N}
\mbox{.\\} \end{equation}
 Using relation (\ref{MLTdchemistry}), we get
\begin{equation}
\label{MLTdT}
\frac{\Delta T}{T}=\frac{\lambda}{2 H_p}(\nabla-\nabla_a)
\left(1+\beta\frac{\chi}{u \lambda}\right)^{-1}
\end{equation}
  where $\nabla=\dpart{\ln T}{\ln p}$ and $H_p^{-1}=-\dpart{\ln p}{r}$.
 We now identify the mixing length of MLT with $\lambda$, so that
 equation (\ref{MLTdT}) becomes the MLT excess temperature equation.
 The density difference can be written in terms of temperature and
 abundance differences if we use the thermodynamical relation
\begin{equation}
\Delta \ln \rho=-\delta \Delta \ln T+{\bmu}''.\Delta{\bf \ln N},
\end{equation}
 where $\delta=-\dth{\ln \rho}{\ln T}{p,{\bf N}}$ and ${\bf
\bmu}''=\dth{\ln \rho}{\bf \ln N}{p,T}$. We obtain the convective
velocity $u$ by substituting this expression into equation
(\ref{MLTdmomentum}) and using relations (\ref{MLTdchemistry}) and
(\ref{MLTdT})

\begin{equation}
\label{cubic}
u^3+u_0u^2
+u_1^2
[\delta (-\nabla+\nabla_a)+\bmu''.{\bf \bnabla_N}]u
+u_0u_1^2\bmu''.{\bf \bnabla_N}=0,
\end{equation}
  where $u_0=\beta\frac{\chi}{\lambda}$,
  $u_1=\sqrt{\frac{p}{\rho}}\frac{\lambda}{2 H_p}$, and ${\bf
  \bnabla_N}=\dd{\bf \ln N}{\ln p}$. Note that the dot product
  $\bmu''.{\bf \bnabla_N}$ corresponds to the more familiar $\phi
  \nabla_{\mu}$ where $\mu$ is mean molecular weight, $\phi=\dth{\ln
  \rho}{\ln \mu}{p,T}$ and $\nabla_{\mu}=\dd{\ln \mu}{\ln p}$.  
Solving this cubic equation for the real positive roots (if they
exist) gives the convective velocity which in turn allows to compute
all the $\Delta$ quantities~:
\begin{eqnarray}
 \Delta \ln T&=&\frac{\lambda}{2 H_p}\frac{u}{u+u_0}(\nabla-\nabla_a)\mbox{ and}\\
 \Delta \ln {\bf N}&=&\frac{\lambda}{2 H_p}{\bf \bnabla_N }.
\end{eqnarray}
  We now determine the number of solutions of the cubic (\ref{cubic}).

\subsubsection{Multiplicity of stationary states}

 This cubic may have three real roots or one real root and two complex
conjugate roots according to the sign of its discriminant
(positive or negative).

 The number of real positive roots relies on the respective signs of
 $\delta(\nabla-\nabla_a)-\bmu''.{\bf \bnabla_N}$ (Ledoux criterion) and $
 \bmu''.{\bf \bnabla_N}$ (Rayleigh-Taylor criterion). We use the rule of
Descartes to obtain the number of solutions.

\begin{itemize}
 \item If $\bmu''.{\bf \bnabla_N}<0$ there is one real positive root
 (Rayleigh-Taylor instability).

 \item If $\bmu''.{\bf \bnabla_N}>0$ and
 $\delta(\nabla-\nabla_a)-\bmu''.{\bf \bnabla_N}<0$ there is no real positive
root.

 \item If $\bmu''.{\bf \bnabla_N}>0$ and
 $\delta(\nabla-\nabla_a)-\bmu''.{\bf \bnabla_N}>0$ there is no
 positive root if the cubic discriminant is positive and there are two
 positive roots otherwise (convective case).
\end{itemize} 

  In the convective case, when $u\gg u_0$, the maximum convective velocity
becomes
\begin{equation}
\label{uconv}
u=u_1\sqrt{\delta (\nabla-\nabla_a)-\bmu''.\bnabla_{\bf N}}.
\end{equation}

\subsection{Stability}

  To analyse the stability of the stationary states we need the
  time-dependent equations for the convective properties
\begin{equation}
\frac{\lambda}{2} \dpart{u}{t}=
-u_1^2\frac{2H_p}{\lambda}(-\delta \Delta \ln T+\bmu''.\Delta \ln {\bf N})-u^2
\mbox{,} \end{equation}
\begin{equation}
\label{tdt}
\frac{\lambda}{2} \dpart{\Delta \ln T}{t}=
u\frac{\lambda}{2H_p}(\nabla-\nabla_a)-(u+u_0)\Delta \ln T
\end{equation}
and
\begin{equation}
\frac{\lambda}{2} \dpart{\Delta \ln {\bf N}}{t}=
u(\frac{\lambda}{2H_p}\bnabla_{\bf N}-\Delta \ln {\bf N})
\mbox{.} \end{equation}
  In equation (\ref{tdt}) we assume that the evolution of the mean fluid
is slow compared to the convective time scale 
so that time-derivatives of $c_p$, $\bmu'$ and $p$ can be neglected.

  The linear stability of this system at a stationary point
is determined by the eigenvalues of the matrix 
\begin{equation}
\label{matrix}
 \frac{\lambda}{2}\left( 
\begin{array}{ccc} 
-2u & \delta
 u_1^2\frac{2H_p}{\lambda} & -u_1^2\frac{2H_p}{\lambda}\bmu''\\
\frac{\lambda}{2H_p}(\nabla-\nabla_a)-\Delta \ln T&-(u+u_0)& {\bf 0}\\
\frac{\lambda}{2H_p}\bnabla_{\bf N}-\Delta \ln {\bf N}&{\bf 0}&-u.{\bf I}
\end{array}
\right) .
\end{equation}
  where ${\bf I}$ is the identity $N\times N$ matrix ($N$ is the total
  number of species),$\bnabla_{\bf N}$ and $\Delta \ln {\bf N}$ are
  vertical $N$-component vectors, and $\bmu''$ is a horizontal
  $N$-component vector.  The system is stable when the real part of
  all eigenvalues is negative. It is unstable otherwise.

\subsubsection{Radiative state ($u=0$)}
 
In this case the matrix (\ref{matrix}) reads

\begin{equation}
 \frac{\lambda}{2}\left( 
\begin{array}{ccc} 
 0 & \delta u_1^2\frac{2H_p}{\lambda} &
 -u_1^2\frac{2H_p}{\lambda}\bmu''\\
\frac{\lambda}{2H_p}(\nabla-\nabla_a)&-u_0&{\bf 0}\\
\frac{\lambda}{2H_p}\bnabla_{\bf N}&{\bf 0}&{\bf 0}
\end{array}
\right). 
\end{equation}

Its characteristic polynomial is $P_N(x)=-(-x)^{N-1}P(x)$ where $P$ is
the cubic (\ref{cubic}).
Hence when there exists a convective state (a real positive root
of the cubic $P$) the radiative state must be unstable.
If there is no real positive root to $P$ and the discriminant is positive,
the stability depends on whether the real part of the complex conjugate roots
are negative. The unstable case corresponds to semi-convection, a state that
is intrinsically time-dependent.
In all the remaining cases (negative discriminant and no real positive root)
the radiative state is stable.

\subsubsection{Convective state ($u>0$)}

In this case, the matrix (\ref{matrix}) reads   
\begin{equation}
 \frac{\lambda}{2}\left( 
\begin{array}{ccc} 
-2u  & \delta u_1^2\frac{2H_p}{\lambda} &
 -u_1^2\frac{2H_p}{\lambda}\bmu''\\
-\frac{u_0}{u+u_0}\frac{\lambda}{2H_p}(\nabla-\nabla_a)&-(u_0+u)&{\bf 0}\\
{\bf 0}&{\bf 0}&-u.{\bf I}
\end{array}
\right). 
\end{equation}

When there are two available stationary states with different velocities
$u$ the state with the lower velocity
is always unstable, and the state with the higher velocity is always stable.\\
  If $u\gg u_0$ the eigenvalues are real negative $\{-2u,-u\}$ and
the convective state is stable.\\

\section*{Acknowledgements}
PL would like to thank Dr. Mike Montgomery for fruitful discussions.
We also thank the referee for pertinent remarks.
 This work was funded
through a European Research \& Training Network on Type Ia Supernovae
(HPRN-CT-20002-00303).

\bibliographystyle{mn2e}

\end{document}